\documentclass[twocolumn]{svjour3}         
\usepackage{graphicx}
\usepackage{ifpdf}
\pdfoutput=1
\journalname{to appear in Astrophysics and Space Science}
\begin{document}

\title{Chemical Complexity in Galaxies\thanks{This research is supported
by the U.S. National Science Foundation.}
}


\author{Jean L. Turner         \and
       David S. Meier
}


\institute{J. L. Turner \at
              Department of Physics and Astronomy, UCLA, Los Angeles CA 90095-1547 USA \\
              Tel.: +1-310-825-4305\\
              Fax: +1-310-825-4305\\
              \email{turner@astro.ucla.edu}           
           \and
        D. S. Meier \at
              NRAO, Socorro, NM 87801 USA \\\email{dmeier@nrao.edu}\\
}


\maketitle

\begin{abstract} ALMA will be able to detect a broad spectrum of molecular lines
in galaxies. Current observations indicate that the molecular line
emission from galaxies is remarkably variable, even on kpc
scales. Imaging spectroscopy at resolutions of an arcsecond or better 
will reduce the chemical complexity by allowing regions of physical
conditions to be defined and classified.  
\keywords{galaxies: ISM \and
astrochemistry \and galaxies: individual (IC~342, M~82)}
\end{abstract}

\section{Introduction}
\label{intro}
A rich spectrum of molecular lines is available at millimeter and
submillimeter wavelengths for the study of molecular clouds in galaxies.
A recent single dish survey in the $\lambda=2$~mm band of the
starburst Sc galaxy NGC 253 revealed lines from 25 
\cite{Met06}  
of the over 35 molecular species detected to date in external galaxies
(see contribution by S. Mart\' in in this volume).
Within the reach of ALMA will be thousands of transitions of over a
hundred molecules, bringing a vast set of potential diagnostics of
molecular gas and physical conditions within galaxies.

With such chemical riches comes complexity.  Existing observations of the most
commonly used molecular tracers, such as CO, HCN, H$_2$, and ammonia,
often give conflicting results for the properties of molecular gas in
galaxies.  Mean densities inferred over decaparsec scales of $\rm \sim
10^3~cm^{-3}$ to $\rm \sim 10^6~cm^{-5}$, depending on which
molecular tracer is used.  Temperatures inferred from molecular lines can
also vary, from $\sim10$--900~K. While CO emission, particularly in spiral
arms, appears to be optically thick, as expected from observations
of Galactic clouds, there appears to
be a widespread diffuse and emissive optically thin CO component in
the interarm regions of spiral galaxies, mixed in with the thick gas
\cite{Wet90,Cet02}.  

Spatial resolution can provide key information
to resolve the complexity and confusion in
molecular spectra of galaxies, by isolating areas of
common physical conditions. Arcsecond imaging of molecular line
emission with millimeter arrays gives resolutions comparable to 
individual giant molecular clouds 
in the nearest galaxies.  Dense cloud tracers identify where the
dense cloud cores are
 \cite{D92},  
and their gas properties should be very
different from those of diffuse molecular gas found between spiral arms 
\cite{Wet93}.  
Nuclear activity can affect relative molecular abundances 
\cite{HB97}  
on these sizescales (see the contribution
by S. Garc\'ia-Burillo in this volume).  These small-scale  
variations in molecular emission properties are of interest to
astrophysicists, since these are potential diagnostics of physical
processes such as shocks or irradiation. The variations are also of
interest to astrochemists, since where molecules are found within 
galaxies may provide clues toward solving the mystery of how they form.  
ALMA will be a tool for both astrophysicists and astrochemists, 
with exquisite sensitivity to many different molecules
and transitions.

\begin{figure*}
 \includegraphics[width=0.75\textwidth]{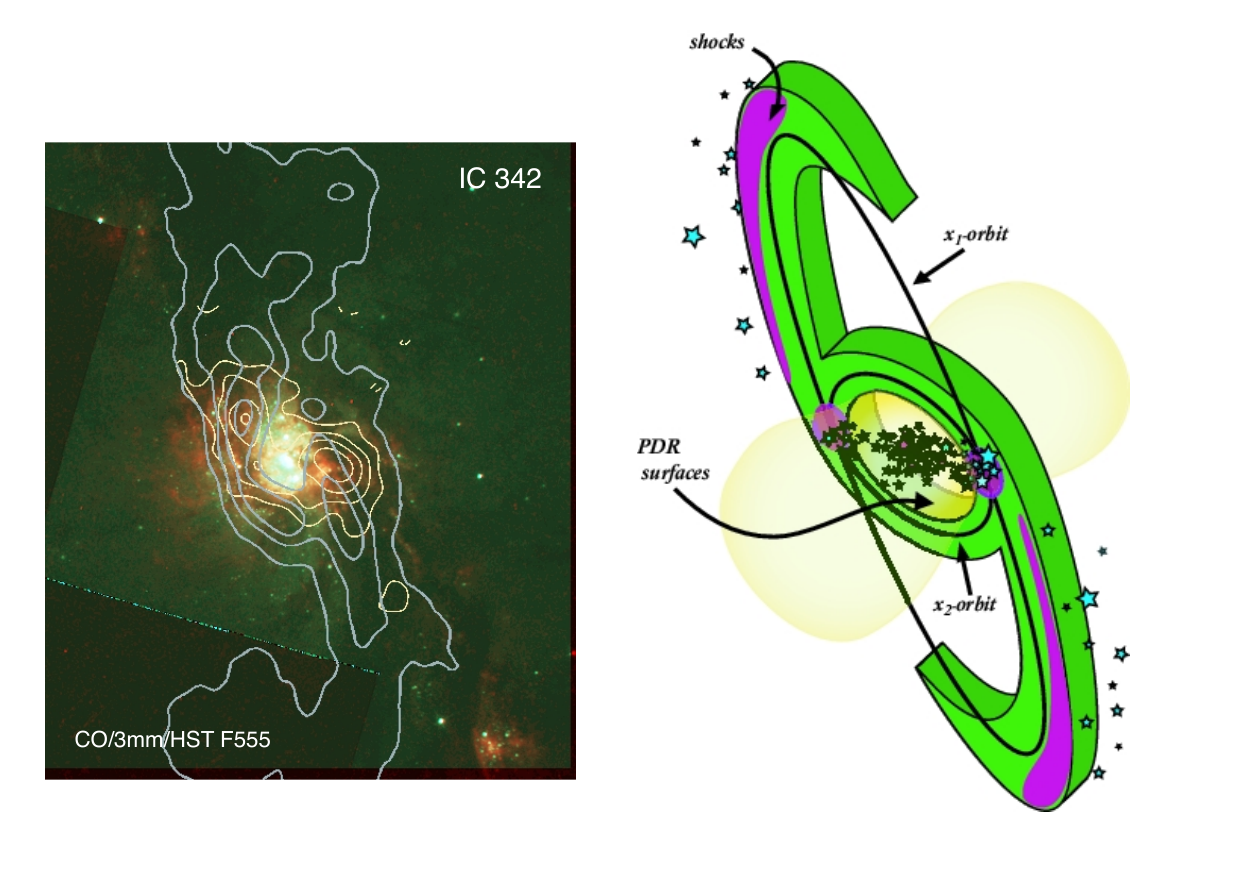}
\centering
\caption{(Left) CO, 3mm continuum, and HST F555 (V) image of the central
300 pc of IC 342. (Right) Schematic of bar orbits and other influences on the
molecular clouds in the center of IC 342.} 
\label{fig1}      
\end{figure*}

An example of the imaging spectroscopy of a nearby normal spiral
galaxy, IC~342, with data from the Owens Valley Millimeter Array, is presented in the
next section.  This example illustrates that with a sufficiently large 
number of spectral lines and
the appropriate statistical analysis, the imaging of the line emission
of many molecules can reduce the complexity of the interpretation of
the physical and chemical processes in the molecular clouds.

\section{Principal Component Analysis of the Center of a Normal Spiral
Galaxy: Imaging Chemistry in IC 342}
\label{sec:1}

IC 342 is a nearby Scd galaxy, nearly face-on, and at a distance of
3~Mpc (1$^{\prime\prime}$=15 pc). An intense episode of nuclear star
formation is indicated by a bright infrared and radio continuum source
of luminosity L$_{IR}\sim 10^8$~L$_\odot$ 
\cite{Bet80,TH83}, 
which is
offset by $\sim$ 50 pc from the dynamical center and nuclear star cluster
\cite{Bet97}. 
Molecular gas in the center of IC~342 forms a
barlike structure 
\cite{Lo84,I90}   
within the central 300 pc.  A
schematic of the stellar $x_1$ and $x_2$ bar orbits and other
dynamical features of this central molecular bar are shown in Figure \ref{fig1} 
\cite{MT05}.  
Streaming of the gas along the bar 
\cite{Lo84} 
leads to
strong shearing motion, correponding to a velocity differential of
50~$\rm km\,s^{-1}$
\cite{TH92} 
across the arms.  The current burst of star
formation is found at the southwestern intersection of the $x_1$ and
$x_2$ bar orbits, where the stellar orbits change from parallel to
perpendicular to the bar 
\cite{MT05,MT01}. 

IC 342 has many faces, depending on which molecules are used to
observe it. Careful analysis of CO and its isotopologues indicate that
the bulk of the molecular gas in the center of IC 342 tends to be
cool, 20~K or less 
\cite{MT01}. 
Dust temperatures measured to
160$\mu$m with the KAO are $\sim$40~K 
\cite{RH84}.  
There is evidence
for an extremely warm component of molecular gas, ranging from
50--150~K as indicated by the lower inversion lines of ammonia,
CO(7--6) and the H$_2$ rotational lines 
\cite{MH86,Het91,Ret02,Met03}.  
Gas temperatures as high as 800--900~K are inferred from the higher
inversion lines of ammonia 
\cite{Met03} 
within the same general region as the
cooler CO gas. However, the velocity of the CO(7--6) emission suggests
that it originates in the clouds to the northeast of the dynamical center
\cite{Het91}, 
a preliminary indication
that CO alone gives an incomplete view of the spatial distribution
of molecular line emission in IC 342.

\begin{figure*}
  \includegraphics[width=0.75\textwidth]{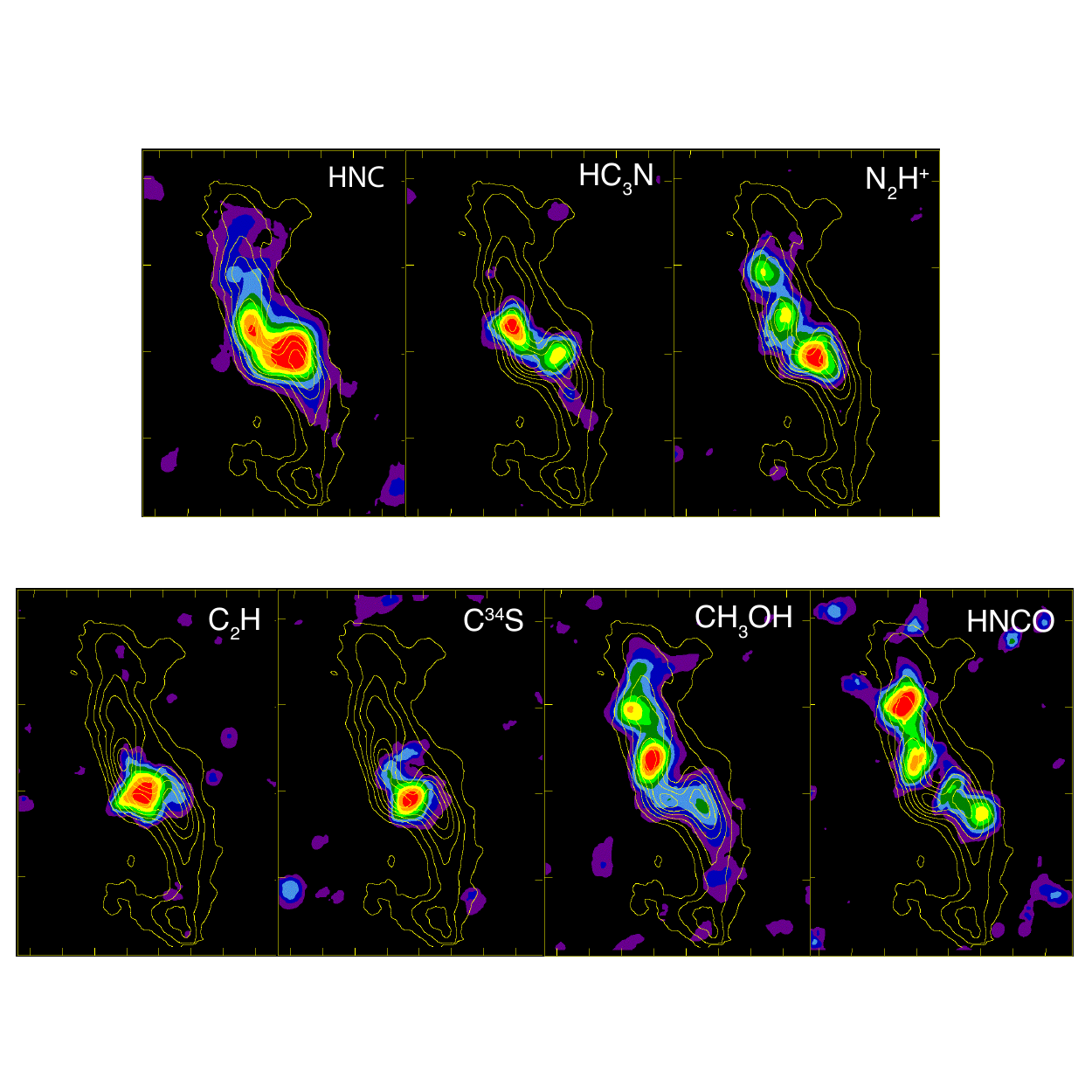}
\centering
\caption{Images of 3mm transitions of molecules in the center of IC 342. Data
from the Owens Valley Millimeter Array at a resolution of 4$^{\prime\prime}$.
 Contours of the 3mm emission
of $^{12}$C$^{16}$O are shown in each panel, showing where the molecular clouds
are found. From \cite{MT05}.} 
\label{fig2}      
\end{figure*}

How can higher resolution and more molecules refine our
understanding of the molecular gas in the center of IC~342?  Images
were made of the central kpc of IC~342 in eight molecules with the
Owens Valley Millimeter Array: C$_2$H, C$^{34}$S, N$_2$H$^+$,
CH$_3$OH, HNCO, HNC, HC$_3$N, and SO by Meier \& Turner 
\cite{MT05}. 
Results were
combined with existing observations of the CO and CO isotopologues of the 2--1
and 1--0 transitions 
\cite{MT01} 
and HCN 
\cite{D92} 
maps. The seven
images (SO was not detected) are shown in Figure \ref{fig2}. 
The 5$^{\prime\prime}$ resolution maps 
reveal a remarkable degree of chemical
variation across the central kpc.  These lines arise from
molecules with similar upper level energies and critical densities;
the differences appear to be due to molecular abundance variations
within the nuclear region 
\cite{MT05}. 

The combination of nearly a dozen different molecular lines with an
image containing $\sim 220$ independent resolution elements is a dataset 
sufficiently large that one can study the statistical correlations among the molecules as
a function of location.  Following the methods of Ungerechts et
al. for the Orion Ridge
 \cite{Uet97}, 
 a principal component analysis was
done of the molecular line intensities across the nucleus of IC~342.
The principal component analysis extracts an unbiased set of
correlations from the data by choosing independent axes that maximize
variance. Strong correlations are apparent in the first two principal
components; a third component seems to indicate weak radial variations.

\begin{figure*}
  \includegraphics[width=0.75\textwidth]{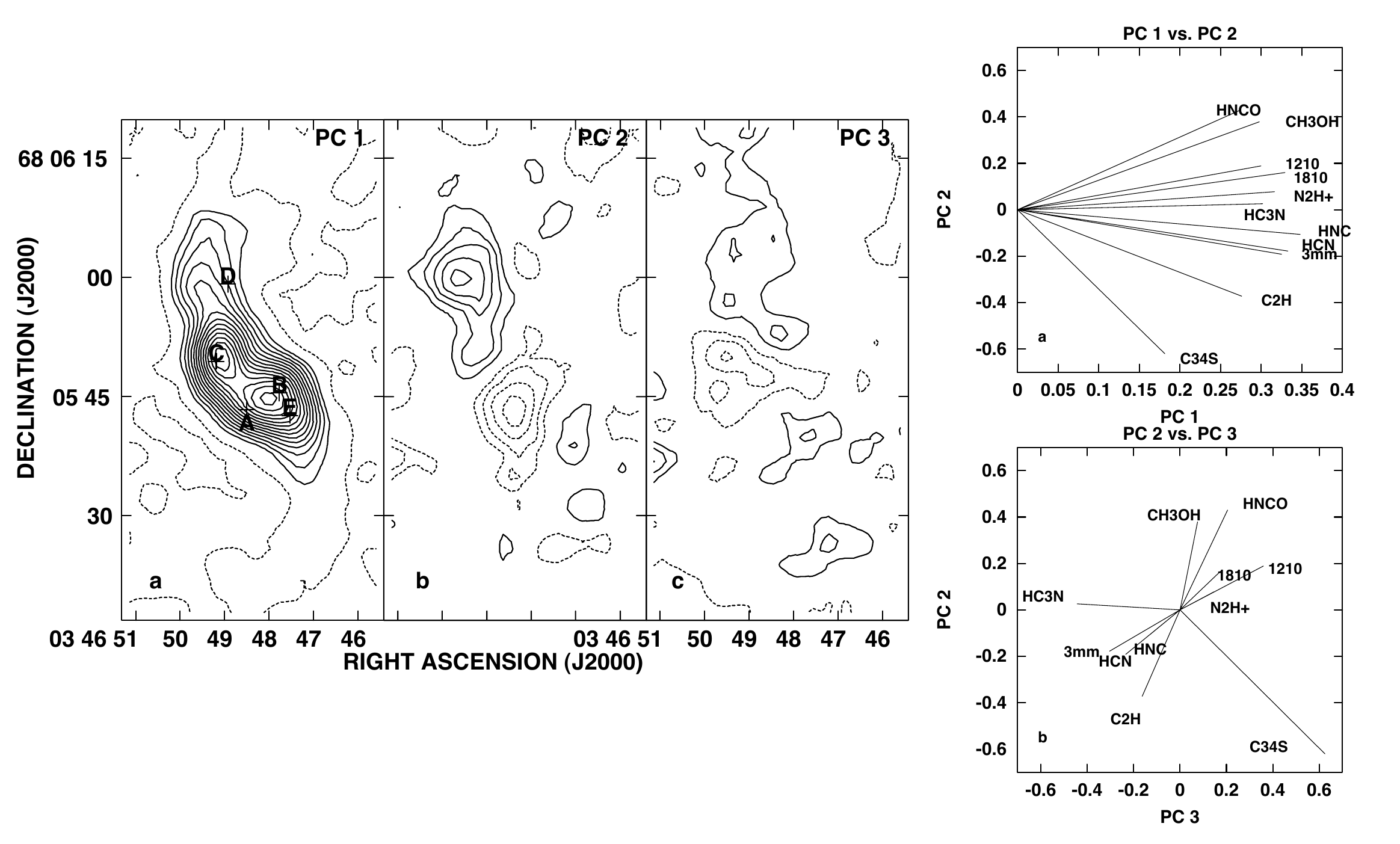}
\centering
\caption{Principal component maps of 3~mm lines in IC~342 and the projections of
the molecules along the principal axes. From \cite{MT05}.
Letters indicate individual molecular clouds identified in higher resolution maps \cite{D92}.}
\label{fig3}       
\end{figure*}

The spatial map of principal component axis 1 is shown in Figure 
\ref{fig3}. With a resemblance to the CO map of Figure \ref{fig1}, 
it represents the best ``average" map for this set of molecules,
which trace a slightly denser gas component than CO. 
This first axis simply tells us that the molecular
lines arise in molecular clouds. More precisely, principal component
axis 1 appears to be the density-weighted mean column density.  The
molecules with the highest projections (i.e., best correlation with
density-weighted mean column density) along PC axis 1 are C$^{18}$O,
N$_2$H$^+$, HNC, and HCN.  In fact the PC 1 map appears to be an
effective average of the C$^{18}$O and HNC maps.  These 3mm lines
appear to be the best general tracers of quiescent clouds and the
overall denser molecular gas component characteristic of the
nuclear region of a spiral galaxy.

Principal component axis 2 distinguishes two further major
correlations, which are roughly anticorrelated here (modulo the
requirement of PC axis 1 that molecules are found in molecular
clouds.) With positive projections along PC axis 2 are the emission of
HNCO and CH$_3$OH (methanol). These molecules are found preferentially
along the molecular bar arms.  Observed Galactic abundances of
methanol are very difficult to produce through gas-phase chemistry, so
methanol is believed to be a tracer of grain chemistry. That methanol
lies along the bar arm in IC~342 suggests that it is produced by the
processing of grain mantles in the shocks along the bar arms. The SiO
image of IC 342 of Usero et al. (2002) looks very much like the CH$_3$OH image 
\cite{Uet06}, 
which would also support this interpretation.  The chemistry of HNCO
has been less certain from Galactic studies; both gas or grain
chemistry have been suggested for it.  Based on the IC~342 image, and
the very good correlation with methanol and SiO, we suggest that it is
also a product of grain chemistry (the excellent correlation of HNCO
and methanol is also seen in the barred galaxy Maffei~2, Meier et
al. in prep.) It is interesting that these putative shock tracers are
found most strongly along the northern bar arm, which also has less
active star formation.

The second correlation evident in PC axis 2 of Figure \ref{fig3} is that of 
C$_2$H and C$^{34}$S. Both of these molecules are restricted to the central 100 pc
of IC~342. The emission from these lines is distinctive in that they
are not found along the bar arms where the other species are found nor
are they coincident with  
the bulk of the youngest embedded star formation. These molecules are
likely to be tracers of highly irradiated molecular gas on the inner
faces of the central ring, which contains a bright nuclear star
cluster, presumably located at the dynamical center of the
galaxy (or a combination of many star clusters, C. Max private
communication).  The fact that these two lines appear to
be more closely associated with the visible nuclear cluster, estimated
to be 60 Myr in age 
\cite{Bet97}, 
than with the IR and radio source of
the current starburst, 
suggests that this gas may be
more closely correlated with a B star population than the young O
stars. A similar result has been found for PAHs in galaxies
\cite{Pet04}. 

Another view of the results of the principal component analysis is the
correlation matrix, shown in Table \ref{tab1}.             
This matrix represents the correlations
between individual pairs of molecules and the 3~mm continuum. Of
particular interest are the unusually low correlations and high
correlations. The spatial anti-correlation of the molecular bar arm
molecules and the nuclear ``PDR" molecules of PC axis 2 described
above is evident in unusually low values of the correlation coefficient. 
The correlation between HNC and HCN is very high, and perhaps this
is not surprising since they are isomers. But the very highest correlation, and it
is remarkably high at 0.94, is between HNC and the 3~mm
continuum. This is closely followed by a correlation between HCN and
3~mm continuum, which at 0.91 is nearly as tight.  The 3~mm continuum
in IC~342 is free-free emission, with little contribution from dust
\cite{MT01}. 
Thus the dense gas tracers HNC and HCN are extremely
well-correlated with free-free emission. This result suggests that the
correlation found by Gao \& Solomon 
\cite{GS04} 
based on global HCN
fluxes (mostly) holds down to the scales of individual giant molecular clouds.

\begin{table*}
\centering
\caption{Correlation matrix for 3~mm molecules in the nucleus of IC~342.}
\label{tab1}       
\begin{tabular}{lcccccccccc}
\hline\noalign{\smallskip}
& $^{12}$CO & C$^{18}$O &3~mm &C$_2$H &C$^{34}$S &CH$_3$OH &HC$_3$N
&HCN &HNC &HNCO \\
\noalign{\smallskip}\hline\noalign{\smallskip}
$^{12}$CO&1.0&$\dots$&$\dots$&$\dots$&$\dots$&$\dots$&$\dots$&$\dots$&$\dots$&$\dots$\\
C$^{18}$O& 0.82& 1.0   &$\dots$&$\dots$&$\dots$&$\dots$&$\dots$&$\dots$&$\dots$&$\dots$ \\
3~mm         & 0.65& 0.76 & 1.0      &$\dots$&$\dots$&$\dots$&$\dots$&$\dots$&$\dots$&$\dots$\\
C$_2$H     & 0.53 & 0.62 & 0.76  & 1.0     & $\dots$ & $\dots$&$\dots$&$\dots$& $\dots$&$\dots$ \\
 C$^{34}$S& 0.38 & 0.39 & 0.48  & 0.50  & 1.0        & $\dots$& $\dots$&$\dots$& $\dots$& $\dots$\\
CH$_3$OH& 0.75 & 0.80& 0.67  & 0.49  & 0.21      & 1.0       &$\dots$&$\dots$&$\dots$&$\dots$  \\
HC$_3$N  &0.60 & 0.71  & 0.85  & 0.57  & 0.30      & 0.68    & 1.0       &$\dots$& $\dots$& $\dots$\\
HCN           &0.65 & 0.75  & 0.90  & 0.74  & 0.49      & 0.64    & 0.77      & 1.0     & $\dots$ & $\dots$\\
HNC           &0.76 & 0.85  & 0.94  & 0.79  & 0.49      & 0.72    & 0.81      & 0.91   & 1.0       & $\dots$ \\
HNCO        &0.67 & 0.75  & 0.58  & 0.42  & 0.19      & 0.78    & 0.57      & 0.57   & 0.66    & 1.0  \\
N$_2$H$^+$&0.73& 0.81& 0.75 & 0.59  & 0.42      & 0.71    & 0.68      & 0.74   & 0.82   & 0.69   \\
\noalign{\smallskip}\hline
\end{tabular}
\end{table*}

While it is encouraging that the ability to resolve features such as
bar arms, orbital intersections, and nuclear star cluster appears to
simplify the chemical analysis, there is still more to learn about the
molecular clouds in 
IC~342. The transitions here all trace the cool and
relatively dense molecular gas
component. While there are indications that the warmer molecular 
cloud component at 50-150~K traced by ammonia, H$_2$,
and upper J levels of CO, and the hot component above 500~K traced by
the upper ammonia inversion levels, are found in different
parts of the nucleus than the lower energy gas traced by CO
\cite{Het91,Het90}, 
the hot molecular gas chemistry remains as yet relatively unexplored in the
imaging domain.

\section{A Different Kind of Chemistry: M82}

A different situation is represented by the dwarf starburst
galaxy, M82. At L$_{IR}\sim 3-4\times 10^9~$L$_\odot$, M82 is one of
the most energetic local starbursts. Since it is a dwarf galaxy, 
processes associated with the starburst might be expected to dominate
the physical processes in the molecular gas of this galaxy, including
the chemistry. These processes would include high
radiation fields and shocks associated with the many 
supernova remnants within the nucleus.

\begin{figure*}
  \includegraphics[width=0.75\textwidth]{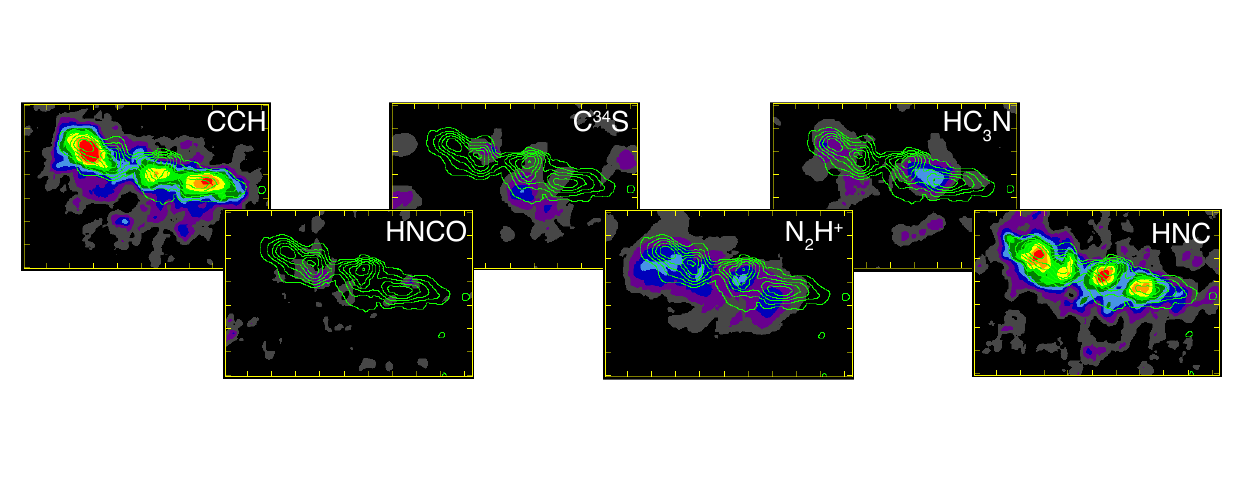}
\centering
\caption{Imaging of 3~mm molecular lines in M82, made using the 
Berkeley-Illinois-Maryland Association Array (Meier and Turner in prep.)}
\label{fig4}       
\end{figure*}

In Figure \ref{fig4} are shown the same set of molecules mapped in M82 
as were mapped in IC~342, observed with the Berkeley-Illinois-Maryland Assocation
array.  Emission from C$_2$H is particularly bright in M82, which
reflects chemistry in the presence of strong radiation fields.  
 HNC is also bright. Based on our analysis of IC~342,
 we infer this is because HNC traces the regions of free-free
emission within the starburst, which are extensive in M82. 
Methanol and HNCO, tracers of grain chemistry, are
relatively weak in M82 compared to the dense gas tracer molecules and
the PDR molecules. This suggests that either the many supernova
remnants observed in M82 either do not have as great an impact on the
molecular gas as do the bar arms in IC~342 or that the harsh interstellar 
radiation field has chemically processed these species even further.
This set of images is consistent with the
characterization of M82 as a ``giant PDR" in terms of its chemistry
\cite{Met91,GBet02}. 

\section{Summary}
\label{sec:4}

High spatial resolution observations of millimeter-wave emission
from molecules indicates that the chemical structure of nearby
galaxies is variable and complex, but that it is possible to spatially
isolate clouds with particular characteristics, such as clouds
experiencing shocks within spiral arms, or clouds near
high radiation fields.  However, this work is at
the limits of the feasibility of today's interferometers, requiring
long integration times and multiple spectral tunings.  ALMA will
revolutionize the study of chemistry in external galaxies due
to its dramatic increase in resolution, sensitivity, and instantaneous
bandwidth.  At Band 6, ALMA will be able to achieve similar
sensitivities to the current IC 342 dataset ($\sim$ 20 mK) in a 
few minutes over 8 GHz simultaneous bandwidth, with six times 
the spatial resolution.  With this degree of improvement, mapping 
significant fractions of nearby galaxies at resolutions of a few pc in an 
extensive 
set of molecular species becomes possible, permitting the study 
of astrochemistry in different local galactic environments in 
much the fashion as large scale surveys do in our Galaxy today.

Pushing out beyond the local neighborhood of galaxies, 
ALMA will be able to detect
IC~342-like GMCs in the brighter lines (eg. HNC and CH$_{3}$OH, 
\cite{Aet02}; 
see contribution by S. Aalto in this volume) out to %
75 Mpc in 1 hour on source; the fainter lines to Virgo distances at
least.  Arp 220-like systems will be detectable in species like HNC
and HC$_{3}$N to $z \sim 0.1$ 
 in an 8 hr track.  This will open up an array of galaxy types---dwarfs, 
 ULIRGs, mergers, and potentially even
ellipticals---to astrochemical scrutiny.  The recent detections of
HCN(3-2) and HCO$^{+}$(1-0) towards the Cloverleaf galaxy at $z =
2.56$ 
 \cite{Set03,Ret06}  
demonstrate that for
the most luminous systems chemistry is within reach across the
observable universe.  


\bibliographystyle{spmpsci}      


\end{document}